\documentclass[10pt,prd,nofootinbib,preprint]{revtex4}
\usepackage[a4paper, total={7in, 10.5in}]{geometry}
\usepackage{bm}
\usepackage[T1]{fontenc}
\usepackage[utf8]{inputenc}
\usepackage{color}
\usepackage{xcolor}
\usepackage{float}
\usepackage{booktabs}
\usepackage{multirow}
\usepackage{amsmath}
\usepackage{graphicx}
\usepackage{esint}
\usepackage{epsfig}
\usepackage{caption}
\usepackage{subcaption}
\usepackage{hyperref}
\usepackage{cancel}
\usepackage{array}
\usepackage{multirow}
\usepackage{makecell}

\usepackage{slashed}
\usepackage{appendix}
\usepackage{tabularx}
\usepackage{pdfpages}
\usepackage{epstopdf}
\usepackage{siunitx}
\usepackage{soul}
\usepackage{setspace}
\makeatletter
\DeclareFontEncoding{LGR}{}{}

\ProvideTextCommand{\~}{LGR}[1]{\char126#1}

\newcommand{\lyxmathsym}[1]{\ifmmode\begingroup\def\b@ld{bold}
  \text{\ifx\math@version\b@ld\bfseries\fi#1}\endgroup\else#1\fi}

\@ifundefined{textcolor}{}
{%
\definecolor{BLACK}{gray}{0}
 \definecolor{WHITE}{gray}{1}
 \definecolor{RED}{rgb}{1,0,0}
 \definecolor{GREEN}{rgb}{0,1,0}
 \definecolor{BLUE}{rgb}{0,0,1}
 \definecolor{CYAN}{cmyk}{1,0,0,0}
 \definecolor{MAGENTA}{cmyk}{0,1,0,0}
 \definecolor{YELLOW}{cmyk}{0,0,1,0}
}

\usepackage[T1]{fontenc}
\begin{document}
\title{Improved $B\to K^*_0\left(1430\right)$ transition form factors at next-to-leading order in light cone sum rules}
\author{Arslan Sikandar$^{1}$\footnote[1]{asikandar@phys.qau.edu.pk}, M. Jamil Aslam$^{1}$\footnote[2]{jamil@qau.edu.pk}\vspace{0.3cm}} %%

\affiliation{ 
$^1$ Physics Department, Quaid-i-Azam University, 45320, Islamabad, Pakistan\vspace{0.5cm}} %%
\begin{abstract}
We present an improved determination of the $B\to K_0^*(1430)$
transition form factors using light-cone sum rules (LCSR) with an expansion
near the light cone. We include the one-loop radiative corrections to
the twist-2 and asymptotic twist-3 distribution amplitudes of the
scalar meson and perform the calculation consistently in the
$\overline{\mathrm{MS}}$ scheme. It allowed us to quantify the impact of radiative corrections on
the form factors for each twist. A daughter sum rule is used from the $B\to K_0^*(1430)$ LCSR to ascertain Borel mass and threshold parameter simultaneously. We find that the dominant theoretical uncertainties
arise from the $K_0^*(1430)$ distribution amplitudes and its decay
constant. Our results are presented in a form that facilitates the
incorporation of future improvements in these nonperturbative inputs,
including those from lattice QCD and experimental measurements, and
can be readily extended to other scalar mesons.

\end{abstract}
\maketitle

\section{Introduction}\label{sed: Introduction}
The Standard Model (SM) of particle physics, proposed by Glashow, Salam and Weinberg, provides a remarkably successful description of the fundamental particles and their interactions via the electromagnetic, weak, and strong forces. Its predictions have been confirmed with high precision in numerous experiments, crowned by the discovery of the Higgs boson at the Large Hadron Collider (LHC). Nevertheless, rare processes involving heavy-flavor hadrons, particularly $B$-meson decays, offer sensitive probes of the SM and potential windows to physics beyond it. Supported by the enormous experimental data, $B$-meson decays have made a great contribution to the precise testing of the SM, where one of the most important tasks is testing the unitarity of the Cabibbo-Kobayashi-Maskawa (CKM) matrix \cite{Cabibbo:1963yz, Kobayashi:1973fv}.

Semileptonic $B$-meson decays, involving both hadronic and leptonic final states, provide a clean environment for phenomenological studies, as the hadronic and leptonic sectors interact only weakly with each other. This feature significantly reduces theoretical uncertainties. The decay amplitudes involving hadrons are governed by hadronic matrix elements, which can be parametrized in terms of form factors. Among these processes, flavor-changing neutral current (FCNC) transitions have attracted considerable attention due to the observed flavor anomalies. In the SM, FCNC processes are forbidden at tree level and are suppressed by the Glashow–Iliopoulos–Maiani (GIM) mechanism~\cite{Glashow:1970gm}, making them highly sensitive probes of potential new physics (NP). The wealth of experimental data has triggered a lot of theoretical research on $B\to P, V$, where $P$ and $V$ correspond to pseudo-scalar $\left(\pi,\; K\right)$ and vector $\left(\rho,\; K^*\right)$ mesons, respectively, see e.g.,~\cite{Tian:2024ubt,Altmannshofer:2024kxb,Becirevic:2024iyi,Allwicher:2024ncl,Dash:2024crn,Wang:2024prt,Hati:2024ppg,Kim:2024tsm,Andersson:2024nam,Buras:2024ewl,Rosauro-Alcaraz:2024mvx,Karmakar:2024gla,Marzocca:2024hua,Bolton:2024egx,He:2024iju,Chen:2024cll,Hou:2024vyw,Li:2024thq,Gabrielli:2024wys,Loparco:2024olo,Chen:2024jlj,Ho:2024cwk,Fridell:2023ssf,McKeen:2023uzo,Altmannshofer:2023hkn,Datta:2023iln,Berezhnoy:2023rxx,He:2023bnk,Amhis:2023mpj,Abdughani:2023dlr,Allwicher:2023xba,Bause:2023mfe,Becirevic:2023aov,Athron:2023hmz}. However, there are still very few investigations on the $B$- meson decaying into an orbital excited state or other excited states; therefore, it is important to study the complementary $B\to S$, where $S$ is a scalar meson which in our case corresponds to $K_0^*\left(1430\right)$.

Compared to the $P$ and $V$ mesons, the internal structure of scalar mesons remains unsettled. The scalar mesons below $1.7~\mathrm{GeV}$ can be classified into two nonets, lying below and above $1~\mathrm{GeV}$, respectively~\cite{Close:2000yk,Close:2002zu,Close:2005vf}. The higher-mass nonet is generally interpreted as consisting of $q\bar{q}$ states; however, it is still debated whether these correspond to the lowest-lying $p$-wave states or to their first radial excitations~\cite{Cheng:2005nb,Lu:2006fr,Du:2004ki,Chen:2021oul,Cheng:2006hu,Yuan:2011xz,Fariborz:2015era}. 
In contrast, there is no consensus on the structure of the scalar-meson nonet below $1~\mathrm{GeV}$. Proposed interpretations include conventional $q\bar{q}$ configurations, tetraquark states, meson--meson molecular states, and, in some cases, glueball- admixtures~\cite{Cheng:2005nb,Cheng:2019tgh,Cheng:2023knr,Weinstein:1982gc,Jaffe:1976ig,Jaffe:1976ih,Amsler:1995tu,Amsler:1995td,Amsler:2002ey}.

The study of semileptonic $B \to S$ decays requires reliable determinations of the $B \to S$ transition form factors. These quantities are nonperturbative in nature and must be evaluated using QCD-based methods. A variety of approaches have been developed for this purpose, including the light-front approach~\cite{Zhang:1994hg, Cheung:1995ub, Choi:1999nu, Chen:2007na}, QCD sum rules (QCDSR)~\cite{Yang:2005bv, Aliev:2007rq, Shifman:1978by}, light-cone sum rules (LCSR)~\cite{Wang:2008da, Sun:2010nv, Han:2013zg, Wang:2014vra, Huang:2022xny, Khosravi:2022fzo, Braun:1988qv, Chernyak:1990ag}, and perturbative QCD (pQCD)~\cite{Li:2008tk, Keum:2000ph, Lu:2000em}. 
Recently, Han \textit{et al.}~\cite{Han:2023pgf} computed the next-to-leading-order (NLO) QCD corrections to the $B \to S$ form factors within the framework of QCD LCSR using $B$-meson light-cone distribution amplitudes (DAs). In this formulation, the $B$-meson-to-vacuum correlation functions are factorized into convolutions of short-distance coefficients and light-cone DAs at the one-loop level. The NLO contributions were found to enhance the form factors by about $5\%$ relative to the leading-order (LO) results. The resulting form factors have subsequently been used to analyze observables such as branching ratios and lepton polarization asymmetries in $B \to S \ell \nu_\ell$ decays.

 Furthermore, by considering the scalar meson nonets below and above $1$ GeV as $q\bar{q}$ states, the two different scenarios of the ground state were proposed \cite{Wang:2014vra}. In first case, the scalar meson below $1$ GeV were considered to be the ground state (Case-I) and in the second scenario, the one above $1$GeV were taken as the ground state (Case-II). In both cases, by taking in to account the $\mathcal{O}\left(\alpha_s\right)$ corrections to the twist-2 terms using LCSR,  the form factors for $B\to S$ transitions were calculated by Wang \textit{et al.} \cite{Wang:2014vra}. It was found that the perturbative $\mathcal{O}\left(\alpha_s\right)$ corrections are about $(15 - 35)\%$ and $(5-20)\%$ for the Cases I and II, respectively.  

 The $B$-meson decay to scalar mesons above 1 $\mathrm{GeV}$ are slightly ambiguous when studied in LCSR compared to light pion. One does get the need to expand the DAs near the light-cone to incorporate the mass effects of the scalar mesons. It was first proposed by Ball \textit{et al.} \cite{Ball:1998sk, Ball:2001fp} for $B$ decay to vector and $K$ meson. The principle is --  near the light-cone, the spacetime variable $x^2\neq 0$ and the momentum of outgoing meson is $P^2=m_S^2$ instead of $P^2=0$. The expansion in $x$ is kept at $\mathcal{O}(x^2)$ and it gives added contributions to the DAs. 

  As mentioned above, for $B\to S$ form factors in LCSR approach, Wang \textit{et al.}~\cite{Wang:2014vra} included the NLO corrections to the twist-2 contribution, the twist-3 terms were retained only at LO. Here, we extend the analysis by incorporating the NLO twist-3 corrections in the asymptotic limit of the corresponding DAs. In all the form factors, the NLO effects are significant for twist-2, where as twist-3 contributions are more pronounced in $f^T$. It indicates that these contributions are essential for a consistent NLO treatment and for the quantitatively reliable predictions suitable for comparison with experimental data.
 
The effective threshold $s_0^B$ and the Borel parameter $M^2$ are
determined simultaneously using the daughter sum rule for the $B$-meson mass associated with the $B\to K_0^*(1430)$ LCSR. In
Ref.~\cite{Duplancic:2008ix}, the Gegenbauer moments of pion DAs were constrained by using the experimentally measured $q^2$ shape of
the $B\to\pi$ form factors. Similarly, future experimental measurements of $B\to K_0^*(1430)$ decays could provide useful constraints on the dominant sources of uncertainty in the present analysis, namely the correlated decay constant and Gegenbauer moments of the $K_0^*(1430)$ DAs.
Throughout the calculation, we employ the $b$-quark mass in the
$\overline{\mathrm{MS}}$ scheme, $m_b(\mu)$. For consistency with the
form factor sum rules, which are evaluated including
$\mathcal{O}(\alpha_s)$ corrections in the $\overline{\mathrm{MS}}$
scheme, we determine the $B$-meson decay constant $f_B$ from a
two-point QCDSR at the same perturbative order and in the same
renormalization scheme~\cite{Jamin:2001fw}.

 The remainder of this paper is organized as follows. In Sec.~\ref{sec:Dist-Amplitude}, we present the DAs for scalar mesons using the expansion near the light cone. Section~\ref{sec:LCSR} is devoted to the derivation of the light-cone sum rules for the $B\to S$ transition form factors at leading and next-to-leading order. In this section, we also discuss the cancellation of infrared divergences in the twist-2 and twist-3 contributions. Section~\ref{sec:Numerical-Analysis} begins with the determination of the continuum threshold and Borel parameter and subsequently presents the numerical inputs and the analysis of theoretical uncertainties. We then give our numerical predictions for the form factors and their $r_1\equiv q^2/m_b^2$ dependence. Finally, Sec.~\ref{sec:conlcusion} summarizes the main results and conclusions of this work.

The paper is supplemented by three appendices. Appendix~\ref{App A-RG} collects the formulae for the renormalization-group evolution of the scalar-meson DAs. Appendix~\ref{App B-daugter SR} presents the daughter sum rule used to determine the continuum threshold. Appendix~\ref{App C-daugter SR} contains the two-point sum rule employed for the determination of the $B$-meson decay constant $f_B$.
\section{Distribution Amplitudes of Scalar Mesons}\label{sec:Dist-Amplitude}

%At NLO the correlation function is given as 
%\begin{equation}
 %   \Pi(q^2),(p+q)^2)=\Pi_0(q^2,(p+q)^2)+\frac{\alpha_s C_F}{4\pi}\Pi_1 (q^2,(p+q)^2),
%\end{equation}
For the the leading Fock (two-particle) state, the light-cone DAs of scalar meson are 
\begin{eqnarray}
    \langle S(p)|\bar q_2(z)\gamma_\mu q_1(0)|0\rangle&=&p_\mu\int_0^1 du\ e^{i u p\cdot z}\Phi_{2S}(u,\mu),\notag\\
    \langle S(p)|\bar q_2(z) q_1(0)|0\rangle &=&m_S\int_0^1 du\ e^{iu p\cdot z}\Phi_{3S}^S(u,\mu),\notag\\
    \langle S(p)|\bar q_2(z)\sigma_{\mu\nu} q_1(0)|0\rangle &=&-m_S(p_\mu z_\nu-p_\nu z_\mu)\int_0^1 du\ e^{iu p\cdot z}\Phi_{3S}^\sigma(u,\mu)\;,\label{vac2sca}
\end{eqnarray}
where $z$ and $p$ denote the light-cone position and momentum variables, respectively, satisfying $z^2=p^2=0$. Here, $u$ and $\bar{u}\equiv1-u$ denote the momentum fractions carried by the quark $q_1$ and antiquark $q_2$, respectively. The function $\Phi_{2S}(u,\mu)$ is the twist-2 DA, while $\Phi_{3S}^{s}(u,\mu)$ and $\Phi_{3S}^{\sigma}(u,\mu)$ are twist-3 DAs that are antisymmetric and symmetric, respectively, under the interchange $u\leftrightarrow1-u$ in the $\mathrm{SU}(3)$ limit, as dictated by $G$-parity conservation. The corresponding normalization conditions for these DAs are given by
\begin{eqnarray}
    \int_0^1 du \Phi_{2S}(u,\mu)=f_S, \quad \int_0^1 du \Phi_{3S}^S(u,\mu)=\int_0^1 du \Phi_{3S}^\sigma(u,\mu)=\bar f_S\;.
\end{eqnarray}

The vector decay constant $f_S$ vanishes in the
$\mathrm{SU}(3)$-flavor symmetry limit. Away from this limit, it is
defined by
\begin{equation}
    \langle S(p)|\bar q_2\gamma^\mu q_1|0\rangle
    = f_S p^\mu .
\end{equation}
Using the equations of motion, $f_S$ is related to the scalar decay
constant $\bar f_S$, defined through
\begin{equation}
    \langle S(p)|\bar q_2 q_1|0\rangle
    = m_S \bar f_S ,
\end{equation}
according to
\begin{equation}
    \bar f_S
    = \frac{m_S}{m_2(\mu)-m_1(\mu)}\,f_S
    \equiv \mu_S^{-1} f_S ,
\end{equation}
where $m_{1,2}(\mu)$ are the $\overline{\mathrm{MS}}$ running quark
masses at the renormalization scale $\mu$. Although $f_S$ vanishes in the $\mathrm{SU}(3)$-flavor limit, $\bar f_S$ remains finite because $f_S$ is proportional to $m_2(\mu)-m_1(\mu)$ in this limit. For scalar mesons
containing a strange quark, such as the $K_0^*(1430)$, the corresponding
flavor-symmetry breaking is therefore reflected in the magnitude of
$\bar f_S$.  This magnitude is then projected on the DAs as well. Though, twist-3 DA is suppressed  compared to twist-2 in the operator product expansion (OPE), it is numerically large due to chiral enhancement. 

Since the scalar meson we are interested here, \textit{i.e.}, $K_0^*\left(1430\right)$ has the mass around $1.5$ GeV, one may consider expanding the DAs near the light-cone in terms of two new variables $x$ and $P$ such that $P^2=m_S^2$ and $x^2\neq 0$. The new variables are defined as follows \cite{Ball:1998sk, Ball:2001fp}:
\begin{align}
    z_\mu &= x_\mu -\frac{x\cdot P-\sqrt{(x\cdot P)^2-m_S^2 x^2}}{m_
    S^2}P_\mu\;,\label{z2X}\\
    p_\mu &= P_\mu -\frac{m_S^2}{2(p\cdot z)}z_\mu=\left(1+\frac{x\cdot P-\sqrt{(x\cdot P)^2-m_S^2 x^2}}{m_S^2}\right)P_\mu-\frac{m_S^2}{2(p\cdot z)}x_\mu\;, \label{p2P}
\end{align}
where $p\cdot z=P\cdot z=\sqrt{(x\cdot P)^2-m_S^2 x^2}$. The above definitions are useful for constructing the corresponding
DAs. The light-cone expansion of the
$b$-quark propagator is given by \cite{Balitsky:1987bk}
\begin{align}
    \langle0|b^i_\alpha(z)\bar b^j_\beta(0)|0\rangle
    &=-i\int \frac{d^4k}{(2\pi)^4}e^{-ik\cdot z}
    \left[
    \delta^{ij}\frac{\slashed{k}+m_b}{m_b^2-k^2}
    \right. \notag\\
    &\left.\quad
    +g_s\int_0^1dv\,G^{\mu\nu a}(vz)
    \left(\frac{\lambda^a}{2}\right)^{ij}
    \left(
    \frac{\slashed{k}+m_b}
    {2(m_b^2-k^2)^2}\sigma_{\mu\nu}
    +\frac{v z_\mu\gamma_\nu}{m_b^2-k^2}
    \right)
    \right]_{\alpha\beta},
    \label{bprop}
\end{align}
where $i,j$ are quark color indices. In the present analysis, we neglect
three-particle contributions involving an additional gluon field. Their
relative importance can be assessed using the equations of motion,
which relate the two-particle and three-particle DAs. The three-particle twist-3 LCDA of \(K_0^*(1430)\) has been estimated from QCD sum rules \cite{Han:2013zg}, and it was found to give less than 1\%  correction to the two-particle twist-3 contribution.
 
The two-particle DAs are defined through
the decomposition of the bilocal vacuum-to-scalar-meson matrix element
in Eq.~(\ref{vac2sca})
\begin{equation}
    \langle S(p)|\bar q_{2\alpha}^i(z)q_{1\beta}^j(0)|0\rangle
    =
    \frac{\delta^{ij}}{12}
    \int_0^1 du\,e^{iu p\cdot z}
    \left\{
    \slashed{p}\Phi_{2S}(u)
    +m_S\Phi_{3S}^s(u)
    -\frac{1}{6}m_S\sigma_{\mu\nu}p^\mu z^\nu
    \Phi_{3S}^\sigma(u)
    \right\}_{\beta\alpha}.
    \label{DA}
\end{equation}
Using the definitions in Eqs.~(\ref{z2X}) and~(\ref{p2P}), the
bilocal matrix element can be expanded near the light cone and,
retaining terms through $\mathcal{O}(x^2)$, takes the form
\begin{eqnarray}
   \langle S(P)|\bar q_{2\alpha}^i(x)q_{1\beta}^j(0)|0\rangle
   &=&
   \frac{\delta^{ij}}{12}\int_0^1du\,e^{iuP\cdot x}
   \Bigg\{
   \left(
   \frac{\slashed{P}}{12}
   -\frac{m_S^2\slashed{x}}{24(P\cdot x)}
   -\frac{m_S^2x^2\slashed{P}}
   {48(P\cdot x)^2}
   \right)\Phi_{2S}(u)
   \notag\\
   &&\qquad
   +\frac{m_S}{12}\Phi_{3S}^S(u)
   +\frac{i m_S}{72}
   \left(P\cdot x-\slashed{x}\slashed{P}\right)
   \Phi_{3S}^\sigma(u)
   \Bigg\}_{\beta\alpha}.
   \label{Smatrixele}
\end{eqnarray}

To transform the coordinate-space expression to momentum space, we
use the substitutions
\begin{align}
x^\mu
&\rightarrow
-i\frac{\partial}{\partial(uP)_\mu},
\nonumber\\
\frac{\phi(u)}{P\cdot x}
&\rightarrow
-i\int_0^u dv\,\frac{\phi(v)}{v}
\equiv -i\phi^{(1)}(u),
\nonumber\\
\frac{\phi(u)}{(P\cdot x)^2}
&\rightarrow
(-i)^2\int_0^u dv\int_0^v dw\,\frac{\phi(w)}{w}
\equiv -\phi^{(2)}(u).
\label{coordinate_replacement}
\end{align}
In obtaining these expressions, integrations by parts have been
performed and the boundary terms have been assumed to vanish,
$\phi(0)=\phi(1)=0$.

The scalar-meson DAs can be expanded in terms of Gegenbauer polynomials with increasing conformal spin:
\begin{align}
\Phi_{2S}(u,\mu)
&=
\bar f_S\,6u\bar u
\left[
B_0(\mu)
+\sum_{m=1}^{\infty}
B_m(\mu)C_m^{3/2}(2u-1)
\right],
\nonumber\\
\Phi_{3S}^{S}(u,\mu)
&=
\bar f_S
\left[
1+\sum_{m=1}^{\infty}
a_m(\mu)C_m^{1/2}(2u-1)
\right],
\nonumber\\
\Phi_{3S}^{\sigma}(u,\mu)
&=
\bar f_S\,6u\bar u
\left[
1+\sum_{m=1}^{\infty}
b_m(\mu)C_m^{3/2}(2u-1)
\right],
\label{DAs}
\end{align}
where $\bar u\equiv1-u$ and
\begin{equation}
B_0(\mu)=\mu_S^{-1}=
\frac{m_2(\mu)-m_1(\mu)}{m_S}.
\end{equation}

\section{Light Cone Sum Rules}\label{sec:LCSR}

The $B\to S$ transition form-factors are parameterized as:
\begin{align}
\langle S(P) | \bar q \gamma_\mu\gamma_5 b | B(P+q) \rangle
&=-
i\left[2 f^{+}(q^2)\, P_\mu
+ \left( f^{+}(q^2) + f^{-}(q^2) \right) q_\mu\right]\;,\notag\\
\langle S(P) |
\bar q\, \sigma_{\mu\nu}\gamma_5 q^\nu b
| B(P+q) \rangle
&=
- \frac{f_{T}(q^2)}{m_B + m_{S}}
\left[
2 q^2 P_\mu - (m_B^2-m_S^2-q^2)\, q_\mu
\right]\;,\notag\\
f^{0}(q^2)
&=
f^{+}(q^2)
+
\frac{q^2}{m_B^2 - m_{S}^2}
\, f^{-}(q^2)\;.\label{ffdef}
\end{align}
%The vacuum-to-scalar meson correlation function %to obtain the LCSR for the form-factors is 
%\begin{equation}
    %\Pi_{\Gamma_{\mu}}(p,q) = i\int d^4x\ e^{i q\cdot x}\langle S(p)|T\left\{\bar q(z)\Gamma_\mu b(z),\ m_b \bar b(0)i\gamma_5d(0)\right\}|0\rangle\;,\label{corrfn}
%\end{equation}
%where the invariant amplitudes for the axial-vector and axial-tensor current can be conveniently written as
%\begin{align}
    %\Pi_{\gamma_\mu\gamma_5}(P,q)&=F(q^2,(P+q)^2)P_\mu+ \widetilde{F}(q^2,(P+q)^2)q_\mu\;,\notag\\
    %\Pi_{\sigma_{\mu\nu}\gamma_5}(P,q)&=F^T(q^2,(P+q)^2)\left(2P_\mu q^2-(m_B^2-m_S^2-q^2)q_\mu\right)\;.\label{invamp}
%\end{align}
To derive the LCSRs for the $B\to S$ transition
form factors, we consider the vacuum-to-scalar-meson correlation
function
\begin{equation}
    \Pi_{\Gamma_\mu}(P,q)
    =
    i\int d^4x\,e^{iq\cdot x}
    \langle S(P)|
    \mathcal{T}\left\{
    \bar q(x)\Gamma_\mu b(x),
    m_b\bar b(0)i\gamma_5 d(0)
    \right\}|0\rangle ,
    \label{corrfn}
\end{equation}
where $P$ is the momentum of the scalar meson and $q$ is the momentum
transferred by the transition current. The Lorentz decomposition of
the correlation function for the axial-vector and axial-tensor
currents can be written in terms of invariant amplitudes as
\begin{align}
    \Pi_{\gamma_\mu\gamma_5}(P,q)
    &=
    F(q^2,(P+q)^2)P_\mu
    +\widetilde F(q^2,(P+q)^2)q_\mu ,
    \notag\\
    \Pi_{\sigma_{\mu\nu}\gamma_5}(P,q)
    &=
    F^T(q^2,(P+q)^2)
    \left[
    2P_\mu q^2
    -(m_B^2-m_S^2-q^2)q_\mu
    \right]\;.
    \label{invamp}
\end{align}
By inserting a complete set of hadronic $B$-meson states into
Eq.~(\ref{corrfn}) and isolating the ground-state contribution, the
hadronic representation of the correlation function, upon using
Eqs.~(\ref{ffdef}) and~(\ref{invamp}), takes the form
\begin{align}
    F(q^2,(P+q)^2)&=\frac{2m_B^2 f_B f^+(q^2)}{m_B^2-(P+q)^2}+\cdots\label{Invamp1}\\%\int_{s_0^B}^\infty ds \frac{\rho^h_+(s,q^2)}{s-(p+q)^2}p_\mu\,\\
     \widetilde{F}(q^2,(P+q)^2)&=\frac{m_B^2 f_B (f^+(q^2)+f^-(q^2))}{m_B^2-(P+q)^2}+\cdots\label{Invamp2}\\
     F^T(q^2,(P+q)^2)&=\frac{2m_B^2 f_B f^T(q^2)}{(m_B+m_S)(m_B^2-(P+q)^2)}+\cdots\label{Invamp3}
\end{align}
On the QCD side, the correlation functions in Eq.~(\ref{corrfn}) are
calculated using the operator product expansion (OPE) in the
deep-Euclidean region, $(P+q)^2<0$. Including the perturbative
contributions through $\mathcal{O}(\alpha_s)$, the corresponding
invariant amplitudes take the form
\begin{equation}
    F(q^2,(P+q)^2))=F_{\text{LO}}(q^2,(P+q)^2)+\frac{\alpha_s  C_F}{4\pi}F_{\text{NLO}}(q^2,(P+q)^2)\;,
\end{equation}
here $F_\text{LO}$ is the LO contribution, while $F_{\text{NLO}}$ is the NLO part. The QCD and hadronic representations of the correlation function are
matched using quark--hadron duality, with the higher-state and continuum
contributions modeled above an effective threshold $s_0^B$. Applying the
Borel transformation in the variable $(P+q)^2$, \textit{i.e.},  $
    (P+q)^2 \to M^2$,
suppresses the contributions from higher resonances and improves the
convergence of the sum rule. The resulting LCSRs for the
$B\to S$ transition form factors are given by
\begin{align}
    f^+(q^2)&=\frac{m_b^2 e^\frac{m_B^2}{M^2}}{2m_B^2f_B}\left[F_{\text{LO}}(q^2,M^2,s_0^B)+\frac{\alpha_s C_F}{4\pi}F_{\text{NLO}}(q^2,M^2,s^B_0)\right]\;,\notag\\
    f^+(q^2)+f^-(q^2)&=\frac{m_b^2 e^\frac{m_B^2}{M^2}}{m_B^2f_B}\left[\widetilde{F}_{\text{LO}}(q^2,M^2,s_0^B)+\frac{\alpha_s C_F}{4\pi}\widetilde{F}_{\text{NLO}}(q^2,M^2,s^B_0)\right]\;,\notag\\
    f^T(q^2)&=\frac{m_b(m_B+m_S)e^\frac{m_B^2}{M^2}}{2m_B^2f_B}\left[F_{\text{LO}}^T(q^2,M^2,s_0^B)+\frac{\alpha_s C_F}{4\pi}F^T_{\text{NLO}}(q^2,M^2,s^B_0)\right]\;.\label{FFgeneralrelation}
\end{align}

\subsection{LCSR at LO}
The tree-level two-particle contribution to the correlation function is
obtained by inserting the leading term of the $b$-quark propagator in
Eq.~(\ref{bprop}) and the scalar-meson two-particle DAs from
Eq.~(\ref{Smatrixele}) into Eq.~(\ref{corrfn}). Defining
\begin{equation}
    \Delta(u) \equiv m_b^2-(q+uP)^2,
\end{equation}
the resulting correlation function at $\mathcal{O}(m_S^2/m_b^2)$ takes the form
\begin{align}
    \Pi_{\gamma_\mu\gamma_5}&=P_\mu \int_0^1\text{d}u\left\{\frac{-m_b}{\Delta}\Phi_{2S}(u)+\frac{m_b m_S^2}{\Delta^2}(-u\Phi^{(1)}_{2S}(u))+\frac{2m_S^2 m_b^3}{\Delta^3}\Phi^{(2)}_{2S}(u)+\frac{um_S}{\Delta}\Phi_{3S}^S(u)\right.\nonumber\\
    &\qquad\left.+\frac{m_S}{6\Delta}\left[2+\frac{m_b^2-u^2P^2+q^2}{\Delta}\right]\Phi_{3S}^\sigma(u)\right\}\nonumber\\
    &+q_\mu\int_0^1 \text{d}u\left\{-\frac{m_b m_S^2}{\Delta^2}\Phi^{(1)}_{2S}(u)+\frac{m_S}{\Delta}\Phi_{3S}^S(u)+\frac{m_S}{6u}\left[\frac{1}{\Delta}-\frac{m_b^2+u^2P^2-q^2}{\Delta^2}\right]\Phi_{3S}^\sigma(u)\right\}\;,\label{corrfnvec}\\
    \Pi_{\sigma_{\mu\nu }q^\nu\gamma_5}&=-\left(-q^2P_\mu+(P\cdot q)q_\mu\right)\int_0^1\text{d}u \left\{\frac{1}{\Delta}\Phi_{2S}(u)+m_S^2\left[\frac{2(m_b^2-u^2 P^2)}{\Delta^3}+\frac{1}{\Delta^2}\right]\Phi^{(2)}_{2S}+\frac{m_bm_S}{3\Delta^2}\Phi_{3S}^\sigma(u)\right\}\;.\label{corrfnten}
\end{align}
The terms involving $\Phi_{2S}^{(1)}$ and $\Phi_{2S}^{(2)}$ represent
new contributions arising from the light-cone expansion. Before
performing the Borel transformation, we use the identity
\begin{align}
    \frac{d}{du}\frac{1}{\Delta}
    &=
    \frac{2P\cdot q+2uP^2}{\Delta^2},
    \nonumber\\
    \frac{1}{\Delta^2}
    &=
    \frac{1}{2P\cdot q+2uP^2}
    \frac{d}{du}\frac{1}{\Delta},
    \label{delta_identity}
\end{align}
where $\Delta =m_b^2-(q+uP)^2$. This relation allows the terms
containing $1/\Delta^2$ to be reduced by integration by parts. The
resulting boundary terms vanish because the relevant DAs vanish at the endpoints, $\Phi_{2S}(0)=\Phi_{2S}(1)=0$.
%Similarly,
%\begin{equation}
 %  \frac{1}{\Delta^3}=\frac{1}{(2P\cdot q+2uP^2)^2}\frac{d}{du}\frac{1}{\Delta}
%\end{equation}
This will help to solve the integral while performing the Borel transform:
\begin{align}
\int \text{d}u \Phi(u)\frac{1}{\Delta^2}&=\int \text{d}u \frac{\Phi(u)}{(2uP^2+2P\cdot q)}\frac{\text{d}}{\text{d}u}\frac{1}{\Delta }\;.\label{integral}
\end{align}
Using integration by parts and the endpoint conditions on the
distribution amplitudes, %followed by the Borel transformation
%$(P+q)^2\to M^2$, 
we obtain
\begin{equation}
\begin{aligned}
\int du\,\frac{\Phi(u)}{\Delta^2}
={}&
-\int du\,\frac{1}{\Delta}
\left\{
\frac{1}{m_b^2-q^2+P^2(2u-1)}
\frac{d\Phi(u)}{du}
-\frac{2P^2\Phi(u)}
{\left[m_b^2-q^2+P^2(2u-1)\right]^2}
\right\}.
\end{aligned}
\label{IBP}
\end{equation}
Including the second term of Eq.~(\ref{IBP}) in the correlation functions in Eqs.~(\ref{corrfnvec},\ref{corrfnten}) would generate contributions of $\mathcal{O}(m_S^3/m_b^3)$ and $\mathcal{O}(m_S^4/m_b^4)$, respectively. These contributions are numerically suppressed and can therefore be neglected. We then obtain the following useful relations:
%The second term in Eq.~(\ref{IBP}) is suppressed by $\mathcal{O}(m_S^2/m_b^2)$  and is therefore neglected in our calculation 
%\textcolor{red}{The terms with $\Delta^2$ or $\Delta^3$ in denominator are of the order $\mathcal(m_S^2/m_b^2$) in Eq. (\ref{corrfnvec}-\ref{corrfnten}). We therefore neglect the second term in Eq. (\ref{IBP}) as it will be of the order $\mathcal{O}(m_S^4/m_b^4)$}. 
\begin{align}
    \int du \frac{\Phi(u)}{\Delta^2}&=-\int du \frac{1}{\Delta}\left(\frac{1}{(m_b^2-q^2+m_S^2(2u-1))}\right)\frac{\text{d}}{\text{d}u}\Phi(u)\;,\notag\\
    \int du \frac{\Phi(u)}{\Delta^3}&=\int du \frac{1}{\Delta}\left(\frac{1}{2(m_b^2-q^2+m_S^2(2u-1))^2}\right)\frac{\text{d}^2}{\text{d}u^2}\Phi(u)\;.
\end{align}

The scaling of terms can even be more elegantly captured by defining dimensionless quantities, i.e., $r_1=q^2/m_b^2,$\  $r_2=(P+q)^2/m_b^2,\ \hat{s}_0^B =s_0^B/m_b^2, \hat{M}^2=M^2/m_b^2$ and $\hat{m}_S^2=m_S^2/m_b^2$. The LO contributions $\left(F_{\text{LO}},\widetilde F_{\text{LO}},F_{\text{LO}}^T\right)$ corresponding to the form factors $\left(f^+(q^2),(f^+(q^2)+f^-(q^2)),f^T(q^2)\right)$ are
\begin{align}
    F_{\text{LO}}(r_1,\hat{M}^2,\hat s_0^B)&=\int_{u_0}^1 du \ \exp\left[-\frac{1-\bar u r_1 +u\bar u\hat{m}_S^2}{u\hat{M}^2}\right]\left\{-\frac{\Phi_{2S}(u)}{u}+\frac{\hat m_S}{6}\left(\frac{2\Phi_{3S}^\sigma(u)}{u}-\left(\frac{1+r_1+u^2\hat m_S^2}{1-r_1+\hat m_S^2(2u-1)}\right)\frac{\text{d}}{\text{d}u}\Phi_{3S}^\sigma(u)\right)\right.\nonumber\\
    &\left.+\hat m_S\Phi_{3S}^S(u)+\frac{\hat m_S^2}{(1-r_1 +\hat m_S^2(2u-1))}\frac{\text{d} }{\text{d}u}\Phi^{(1)}_{2S}(u)+\frac{\hat m_S^2}{2(1-r_1 +\hat m_S^2(2u-1))^2}\frac{\text{d}^2 }{\text{d} u^2}\Phi^{(2)}_{2S}(u)\right\}\;,\label{FplusLO}\\
    \widetilde F_{\text{LO}}(r_1,\hat{M}^2,\hat s_0^B)&=\int_{u_0}^1 du \ \exp\left[-\frac{1-\bar u r_1 +u\bar u \hat{m}_S^2}{u\hat{M}^2}\right]\left\{\frac{\hat m_S\Phi_{3S}^S(u)}{u}+\frac{\hat m_S}{6u}\frac{\text{d}}{\text{d}u}\Phi_{3S}^\sigma(u)+\frac{\hat m_S^2}{(1-r_1+\hat m_S^2(2u-1)}\frac{\text{d} }{\text{d}u}\Phi^{(1)}_{2S}(u)\right\}\;, \label{FtildeLO}\\
      F_{\text{LO}}^T(r_1,\hat{M}^2,\hat s_0^B)&=\int_{u_0}^1 du \ \exp\left[-\frac{1-\bar u r_1 +u\bar u \hat{m}_S^2}{u\hat{M}^2}\right]\left\{\frac{\Phi_{2S}}{u}+\frac{\hat m_S}{3(1-r_1+\hat m_S^2(2u-1))}\frac{\text{d}}{\text{d}u}\Phi_{3S}^\sigma(u)\right.\nonumber\\
      &\qquad\left.+\hat m_S^2\left[\frac{1}{2(1-r_1+\hat m_S^2(2u-1))^2}\frac{\text{d}^2 }{\text{d}u^2}\Phi^{(2)}_{2S}(u)-\frac{1}{2(1-r_1+\hat m_S^2(2u-1))}\frac{\text{d} }{\text{d}u}\Phi^{(2)}_{2S}(u)\right]\right\}\;,\label{FTLO}
\end{align}
where
\begin{equation}
u_0=\frac{\sqrt{(\hat{s}_0^B-r_1^2-\hat m_S^2)^2+4\hat m_S^2(1-r_1^2)}-(\hat{s}_0^B-r_1^2-\hat m_S^2)}{2\hat m_S^2}\;,
\end{equation}
which in the limit $m_S\to 0$ is $u_0=\tfrac{\hat s_0^B-r_1}{r_2-r_1}$.
\subsection{LCSR at NLO}
The NLO amplitude is a convolution of the perturbative hard-scattering elements with the DAs:
\begin{align}
   % F_{\text{LO}}(q^2,(p+q)^2)&=\bar f_S\int_0^1du \Phi_{2S}(u)T_0(q^2,(p+q)^2,u),\\ 
    F_{\text{NLO}}(r_1,\hat{M}^2,\hat{s}^B_0)&=\frac{m_b^2 }{\pi}\int_{1}^{\hat{s}_0^B}dr_2\  e^{-r_2/\hat{M}^2}\int_0^1 du \left\lbrace \text{Im}  \text{T}_1(r_1,r_2, u)\Phi_{2S}(u)\right.\notag\\
    &\qquad+\hat{m}_S\left[\text{Im}  \text{T}_1^S (r_1,r_2, u)\Phi_{3S}^{S,\text{asymp}}+\text{Im}  \text{T}_1^\sigma (r_1,r_2, u)\Phi_{3S}^{\sigma,\text{asymp}}\right]\;,\label{FFNLOgeneralrel}
\end{align}
where $\widetilde{F}_{\rm NLO}$ and $F^T_{\rm NLO}$ have analogous
expressions, with the corresponding imaginary parts, and
$\Phi_{3S}^{(S,\sigma),\mathrm{asymp}}$ denote the asymptotic forms of
the corresponding DAs. It is important to mention that in our analysis both the LO and NLO contributions are expressed in
terms of dimensionless quantities, as given in Eqs.~(\ref{FplusLO})
--~(\ref{FFNLOgeneralrel}). In the present calculation, the convolution
is retained up to twist-3. The NLO corrections are evaluated in a
general covariant gauge. Both ultraviolet (UV) and infrared (IR)
divergences are regulated using dimensional regularization, and the
resulting UV divergences are renormalized in the
$\overline{\mathrm{MS}}$ scheme. We employ an anticommuting $\gamma_5$
prescription throughout.
The correlation function in Eq.~(\ref{corrfn}) contains two
unrenormalized currents,
\begin{equation}
    J_5=\bar b\,\gamma_5 d,
    \qquad
    J_A=\bar q\,\gamma_\mu\gamma_5 b,
    \qquad q=s,u,
\end{equation}
where the corresponding renormalized quantities are related to the
bare ones according to
\begin{equation}
    J_5^{\text{bare}}=Z_5(J_5)^r,\qquad
    J_A^{\text{bare}}=Z_A(J_A)^r,\qquad
    m_b^{\text{bare}}=Z_m(m_b)^r,
    \qquad
    \Delta=\frac{1}{\epsilon}-\gamma_E+\ln 4\pi .
\end{equation}
In the $\overline{\mathrm{MS}}$ scheme, the relevant renormalization
constants are
\begin{equation}
    Z_5
    =1+3\Delta\frac{\alpha_s C_F}{4\pi},
    \qquad
    Z_A=1,
    \qquad
    Z_m
    =1-3\Delta\frac{\alpha_s C_F}{4\pi},
    \qquad
    Z_T
    =1+\Delta\frac{\alpha_s C_F}{4\pi}.
\end{equation}
For the axial-vector current considered here, $Z_A=1$ at the order
under consideration. In particular, the renormalization factors
associated with the pseudoscalar current and the $b$-quark mass satisfy
\begin{equation}
    Z_m Z_5=1+\mathcal{O}(\alpha_s^2),
\end{equation}
so that the ultraviolet renormalization of the correlation function
with the combination $m_b J_5 J_A$ is implemented through
\begin{equation}
    m_b^{\text{bare}} J_5^{\text{bare}} J_A^{\text{bare}}
    \;\longrightarrow\;
    Z_m(m_b)^r\,Z_5(J_5)^r\,Z_A(J_A)^r.
\end{equation}
Consequently, for the vector/axial-vector form factors, the UV
renormalization of the one-loop hard amplitudes can be implemented by
expressing the bare $b$-quark mass in terms of the renormalized mass
$(m_b)^r$, together with the corresponding current renormalization
factors. Since
\begin{equation}
    Z_mZ_5Z_A
    =1+\mathcal{O}(\alpha_s^2),
\end{equation}
no additional overall UV counterterm remains at
$\mathcal{O}(\alpha_s)$.

For the tensor current, which is not conserved and consequently
introduces a renormalization-scale dependence into the corresponding
form factors, an additional renormalization factor $Z_T$ is required.
Thus, the tensor form factors receive, in addition to the mass
renormalization, the multiplicative renormalization associated with
the tensor current.

At NLO, six one-loop diagrams \cite{Ball:2001fp,Duplancic:2008ix,Khodjamirian:1997ub} contribute to the correlation function.
After combining the UV-divergent terms from all diagrams for each
twist contribution, the complete UV divergences cancel against the
counterterms generated by the mass and current renormalization. For
the vector/axial-vector form factors, this cancellation follows from
the replacement of the bare mass by the renormalized mass together with
the relation
\begin{equation}
    Z_mZ_5Z_A
    =1+\mathcal{O}(\alpha_s^2).
\end{equation}
For the tensor form factor, the corresponding cancellation additionally
requires the tensor-current renormalization factor $Z_T$. The resulting
hard amplitudes are therefore ultraviolet finite at NLO.

%We are left with IR divergences that need to be carefully taken care of, otherwise our results will not be finite. For the leading twist $\Phi_{\text{2S}}$, the IR divergences arising in the NLO twist-2 hard scattering diagrams, are factorized into the evolution of the twist function using Brodsky-Lapage kernel[lepage Brodsky] analogously to the pion-twist-2 DA since scalar DA is also a flavor singlet quark-antiquark operator.  The IR divergences coming from the remaining twist-3 functions $\Phi_{3S}^{(S,\sigma),\text{asymp}}$ are canceled in their asymptotic limit and when taken in combination. This was first done in (khojimirian) and later by (Blaz) for the pion case. We just extended it for the scalar meson case. A  more interesting case is the $B$ decays to vector mesons for which we have already shown the IR cancellations for all the seven form-factors at order $m_V^2$ that even include mixed DAs (possible due to expansion near light-cone). This work is in progress for further phenomenology. 

The remaining infrared (IR) divergences require careful treatment to
ensure that the resulting form factors are finite. For the leading-twist
distribution amplitude $\Phi_{\text{2S}}$, the IR divergences arising
from the NLO twist-2 hard-scattering diagrams are factorized into the
evolution of the twist-2 DA through the
Brodsky--Lepage kernel \cite{Lepage:1980fj}, in analogy with the standard
treatment of the pion twist-2 DA. This factorization
is applicable because the scalar-meson DA is likewise
defined through a flavor-singlet quark-antiquark operator.

For the remaining twist-3 DAs,
$\Phi_{3S}^{(S,\sigma),\text{asymp}}$, the IR-divergent contributions
cancel in the asymptotic limit when the corresponding terms are
considered in combination. This procedure was first employed in
\cite{Ball:2001fp} and subsequently to the pion case in  \cite{Duplancic:2008ix}, in which the finite corrections are explicitly given.
Here, we extend the same treatment to the scalar-meson case.
A particularly interesting extension concerns $B$-meson decays into
vector mesons. In a separate work \cite{vector:2026}, we demonstrated the cancellation
of IR divergences for all seven form factors at order $m_V^2$, including
contributions from mixed distribution amplitudes, which become accessible
through an expansion around the light cone. The extension of this
framework to further phenomenological applications is currently in
progress.

\section{Numerical Analysis and results}\label{sec:Numerical-Analysis}
To proceed with the numerical calculation of the form factors, it is important to mention that the Borel parameter $M^2$ and the effective threshold $s_0^B$ are
auxiliary parameters rather than physical observables. They are
introduced to suppress contributions from higher resonances and the
continuum, isolate the ground-state $B$-meson contribution, and
improve the convergence of the OPE. Since the
twist-3 contributions are comparable to, or larger than, the twist-2
contributions for scalar mesons, we anticipate that the optimal Borel
window may be shifted toward somewhat larger values of $M^2$.

Several prescriptions have been employed in the literature to
constrain the Borel parameter and the effective threshold. In
Ref.~\cite{Ball:2004ye}, a two-point QCD sum rule for the $B$-meson
decay constant was used to relate its Borel parameter to that entering
the $B\to\pi$ LCSRs through
\begin{equation}
    M_{\rm LCSR}^2
    =
    c_c\,\frac{\overline{M}_{\rm QCD}^2}{\langle u\rangle}\;,
\end{equation}
where $\langle u\rangle$ is determined from a daughter sum rule and
$\overline{M}_{\rm QCD}^2$ denotes the Borel parameter of the
corresponding two-point QCDSR. In contrast, Ref.~\cite{Duplancic:2008ix}
constrained the Borel window directly from the $B\to\pi$ LCSR. The
parameters $s_0^B$ and $M^2$ were simultaneously adjusted such that the
corresponding daughter sum rule reproduced the experimentally known
$B$-meson mass to approximately $1\%$ accuracy. These parameters were
then used, together with the experimentally measured $q^2$ dependence
of $B\to\pi\ell\nu$, to constrain the Gegenbauer moments $a_2$ and
$a_4$.

The use of the daughter sum rule is particularly useful because the
$B$-meson mass is known with high experimental precision. It is obtained
from the ratio of the derivative of the sum rule with respect to
$-1/M^2$ to the sum rule itself. We adopt this strategy to constrain
$s_0^B$ and the appropriate Borel window for the present
$B\to K_0^*(1430)$ analysis. We determine the effective threshold $s_0^B$ and the Borel window by constructing a daughter sum rule Appendix \ref{App B-daugter SR} for the $B$-meson mass from the $B\to K_0^*(1430)$ LCSR. The higher-twist contributions are anticipated to be small, and we require the NLO corrections to remain below $30\%$ for $f_Bf^+(0)$ to ensure a reliable perturbative expansion. This criterion sets the lower bound of the Borel window at $M_{\min}^2=16\;\mathrm{GeV}^2$. The same requirement on the size of the NLO corrections constrains the renormalization scale from below, leading to $\mu_{\text{min}}=2.5\;\mathrm{GeV}$. The upper limit of the Borel window is fixed by requiring the contribution from higher hadronic states to remain below $30\%$, which yields $M_{\max}^2=22\;\mathrm{GeV}^2$. Within the resulting Borel window,
$16\;\mathrm{GeV}^2\leq M^2\leq22\;\mathrm{GeV}^2$, we find that the choice $s_0^B=35\;\mathrm{GeV}^2$ leads to an extracted $B$-meson mass with an uncertainty of approximately $3\%$, using the central values of the Gegenbauer moments of the $K_0^*(1430)$. This uncertainty is somewhat larger than the $1\%$ uncertainty reported in Ref.~\cite{Duplancic:2008ix}. We also find that the $\mathcal{O}(\alpha_s)$ corrections modify the extracted $B$-meson mass by approximately $8\%$ in the present $B\to K_0^*(1430)$ daughter sum rule. We therefore adopt
$s_0^B=35\;\mathrm{GeV}^2$ and $M^2=18\;\mathrm{GeV}^2$ as the effective
threshold and central Borel parameter, respectively, for our NLO
analysis.

For a consistent extraction of the form factors, we additionally employ a
two-particle sum rule (2pSR) for the $B$-meson decay constant $f_B$,
including the $\mathcal{O}(\alpha_s)$ corrections. This is required to
ensure consistency between the renormalization scheme and the
perturbative order used in the determination of $f_B$ and those used
in the form factor sum rules. The 2pSR is evaluated at NLO in the
$\overline{\mathrm{MS}}$ scheme following
Refs.~\cite{Duplancic:2008ix,Bagan:1997bp,Ball:2004ye} and is given in Appendix \ref{App C-daugter SR} for reference. We determine
the corresponding threshold parameter and Borel window from the
daughter sum rule for the $B$-meson mass. For
$\overline{s}_0^B=34-37\;\mathrm{GeV}^2$ and
$\overline{M}^2=4-6\;\mathrm{GeV}^2$, the extracted $B$-meson mass
varies by approximately $0.5\%$ over the allowed parameter range. We
emphasize that the effective threshold parameter in the 2pSR,
$\overline{s}_0^B$, need not coincide with the threshold parameter
$s_0^B$ entering the $B\to S$ form factors sum rule. In particular, we
do not impose the equality of the two threshold parameters, in
contrast to the prescription adopted in \cite{Ball:2004ye}.

The nonperturbative inputs required for the 2pSR include the light-quark
condensate
\begin{equation}
    \langle\bar q q\rangle(2\,\mathrm{GeV})
    =-(0.272)^3\;\mathrm{GeV}^3,
\end{equation}
taken from the lattice/FLAG determination
\cite{FlavourLatticeAveragingGroup:2019iem}, as well as the gluon
condensate
\begin{equation}
    \left\langle\frac{\alpha_s}{\pi}GG\right\rangle
    =0.012^{+0.006}_{-0.012}\;\mathrm{GeV}^4\;,
\end{equation}
and the mixed quark-gluon condensate parameter
\begin{equation}
m_0^2=0.8\pm0.2\;\mathrm{GeV}^2,
\end{equation}
as taken from Ref.~\cite{Gubler:2018ctz}. Using the daughter sum rule,
we obtain the effective threshold
$\overline{s}_0^B=35.6\;\mathrm{GeV}^2$, with the central Borel
parameter $\overline{M}^2=5\;\mathrm{GeV}^2$. At these central values,
the NLO 2pSR yields, $
f_B=211^{+6}_{-7}\;\mathrm{MeV}$.

The $b$-quark mass in the $\overline{\mathrm{MS}}$ scheme is
renormalization-scale dependent, as is the two-loop strong coupling
$\alpha_s(\mu)$ given in Eq.~(\ref{mbmuasmu}). In addition, the
twist-2 distribution amplitude exhibits scale dependence associated
with its renormalization and factorization with the hard-scattering
kernel. The twist-2 and twist-3 distribution amplitudes in
Eq.~(\ref{DAs}), together with the $K_0^*(1430)$ decay constant, are
given at the reference scale $\mu_0=1\;\mathrm{GeV}$ (see
Table~\ref{Table:input}). We evolve these quantities from
$\mu_0=1\;\mathrm{GeV}$ to a common scale $\mu=3\;\mathrm{GeV}$ using
the renormalization-group equations given in Eq.~(\ref{RGmoments}).
The $\overline{\mathrm{MS}}$ $b$-quark mass is similarly evolved from
$m_b(m_b)$ to $m_b(3\;\mathrm{GeV})$. We consistently perform both the
two-point sum rule calculation for the $B$-meson decay constant $f_B$
and the LCSR calculation of the $B\to K_0^*(1430)$ transition form
factors at the same scale, $\mu=3\;\mathrm{GeV}$. Thus, all
perturbative and hadronic quantities entering the calculation are
consistently evaluated in the $\overline{\mathrm{MS}}$ scheme at the
common scale $\mu=3\;\mathrm{GeV}$.

%The $b-$quark mass in $\overline{\text{MS}}-$scheme is scale dependent as well as the 2-loop strong coupling given in Eq. (\ref{mbmuasmu}). There is also scale dependence in the twist-2 distribution amplitude when convoluted with the hard kernel known as the factorization scale. Moreover, the values of twist-2 and twist-3 distribution amplitudes (\ref{DAs}) along with the decay constant of $K_0^*$ are given at 1 GeV (Table. \ref{Table:input}). We take a single scale $\mu=3\text{GeV}$ and the $b-$quark mass is RG evolved from $mb(mb)$ to its value at 3  GeV along with the Gegenbaur moments and decay constant. Their RG evolution is given in Eq. (\ref{RGmoments}). The two-point sum rule calculation for $f_B$ and the LCSR calculation for $B\to K_0^*$ transition form-factors are done at the same scale, i.e., $\mu=3\ \text{GeV}$. The whole calculation is done in $\overline{\text{MS}}$-scheme. The hadronic quantities given at scale 1 GeV are evolved to 3 GeV.

%The results of the form-factor are obtained by substituting the LO relations, Eq. (\ref{FplusLO},\ref{FtildeLO},\ref{FTLO}) and NLO relation (\ref{FFNLOgeneralrel}) in Eq. (\ref{FFgeneralrelation}). The Borel parameter is chosen here  to be $M^2=18  \;\text{GeV}$. 
\begin{table}[t]
\centering

 \begin{tabular}{||c| c||c| c||} 
 \hline
 Parameter & Value & Parameter & Value \\ [0.5ex] 
 \hline\hline
 $m_{B^0}$ & 5.279 GeV & $m_b(m_b)$ & 4.18 GeV \\ 
 $m_d$(3 GeV) & 4.2$\pm$0.4 MeV & $m_s$(3 GeV) & 84$\pm$ 0.5 MeV\\
$\overline{f}_{K_0^*}(1\text{ GeV})$ & 445$\pm$50 MeV & $\alpha_s$(3 GeV) & 0.255 \\
 $B_1$(1 GeV) & $-0.57$$\pm$0.13 & $B_3$(1 GeV)& $-0.42$$\pm$0.22 \\
 $a_1$(1 GeV) & 0.0018$\sim$ 0.0042 & $a_2$(1 GeV) & $-0.33$$\sim $ $-0.025$ \\
 $b_1$(1  \;GeV)&0.0037$\sim$0.0055&$b_2$(1 GeV)&0$\sim$0.15\\ [1ex] 
 \hline
 \end{tabular}
 \caption{The input values for the quark masses are taken from Ref.~\cite{PDG:2026}, while those for the Gegenbauer moments of $K_0^*(1430)$ are taken from Ref.~\cite{Cheng:2005nb}.} 
 \label{Table:input}
\end{table}

At the LO, the form factor $f^+(0)$ at zero momentum
transfer ($q^2=0$) receives contributions from both the twist-2 and twist-3 DAs, namely $\Phi_2(u)$ and $\Phi_3^{(S,\sigma)}(u)$, respectively. The twist-3 contribution is
parametrically suppressed by a factor of order $\hat m_S$. If the
$\hat{m}_S^2$ corrections arising from the near-light-cone expansion
are neglected, the twist-2 and twist-3 contributions account for
approximately $32\%$ and $68\%$ of the total form factor,
respectively. The numerical dominance of the twist-3 contribution over the twist-2 contribution is expected for scalar mesons because the twist-2 DA is normalized by the vector decay constant $f_S$, which vanishes in the SU(3)-flavor limit, whereas the twist-3 DAs are normalized by the scalar decay constant $\bar f_S$, which remains nonzero in this limit. This behavior can be contrasted with the $B\to\pi$ case studied in \cite{Duplancic:2008ix} , where the twist-2 and twist-3 contributions account for approximately $50.5\%$ and $46.7\%$ of the form factor, respectively. Including the $\hat{m}_S^2$ corrections
further enhances the relative importance of the twist-3 contribution
and correspondingly reduces the twist-2 by 5\% for $f^+(0)$ and $\sim 3\%$ for $f^T(0)$. 

The NLO corrections to the form factors are summarized in Table~\ref{Kresult}. We note that
Ref.~\cite{Wang:2014vra} included NLO radiative corrections only for
the twist-2 contribution, whereas the present analysis also includes
the NLO corrections to the asymptotic twist-3 contributions. All three form factors received sizeable  corrections for twist-2 and small corrections to twist-3 except the form-factor $f^T(0)$. At $q^2=0$, the NLO correction to the quantity $f_Bf^+(0)$ is +10\% of the LO while that  for $f_Bf^T(0)$ is $15.8\%$. The corresponding NLO correction to the two-point sum rule for \(f_B\) is approximately \(21\%\) (for LO $f_B=175 \rm MeV$ and for NLO $f_B= 211 \rm MeV$).  Consequently, upon extracting the form factors by dividing the LCSR by the NLO $f_B$, the corrections are reduced to $-9.4\%$ and $-4.2\%$ for $f^+$ and $f_T$, respectively. This demonstrates a substantial cancellation between the radiative corrections to the LCSR and the two-point sum rule of $B-$ decay constant and a far better stability under the change of scale. Contrast it with \cite{Khodjamirian:1997ub} in which the authors found a radiative correction to NLO twist-2 to be 4.7\% while the authors in  \cite{Duplancic:2008ix} found it 7.4\%. The former argued that there may be possibility of complete cancellation but the latter showed that the inclusion of twist-3 results in partial cancellation which we also agree with.

%The NLO corrections to the LO result, including the mass corrections from the near-light-cone expansion, are found to be significant. The total NLO correction to $f^+(0)$ is approximately $9.5\%$ of the LO result. Separately, the twist-2 and twist-3 contributions receive NLO corrections of approximately $4.2\%$ and $19.7\%$, respectively. The relatively large NLO correction to the twist-3 contribution demonstrates the importance of including the twist-3 radiative corrections, which constitutes one of the main improvements of the present analysis. This behavior can be contrasted with the $B\to\pi$ case studied in \cite{Duplancic:2008ix} , where the twist-2 and twist-3 contributions account for approximately $50.5\%$ and $46.7\%$ of the form factor, respectively, while the corresponding NLO corrections were found to be $7.4\%$ for twist-2 and $-4.4\%$ for twist-3. The comparatively larger NLO correction to the twist-3
%contribution in the present $B\to K_0^*(1430)$ analysis is consistent
%with the enhanced scalar-meson twist-3 contribution, which is
%controlled by the chiral parameter $m_S/[m_s(\mu)-m_q(\mu)]$ rather
%than the corresponding pion parameter $\mu_\pi=m_\pi^2/[m_u(\mu)+m_d(\mu)]$. %\textcolor{red}{A similar calculation was also performed in the pole-mass scheme}

\begin{table}[ht]
    \centering
    \begin{tabular}{|c|c|c|c|c|c|c|c|}
    \hline
         form-factor&twist-2(LO)&twist-3(LO)&twist2(NLO)&twist3(NLO)&Total (LO)&Total(NLO)&Difference  \\ \hline
         $f^+(0)$ &0.174 &0.382&0.124&0.378&0.555&0.503&$-9.4$\% \\
         \hline
         $f^+(0)+f^-(0)$&$-0.033$&0.189&$-0.024$&0.205&0.163&0.181&$-11$\%\\
         \hline
          $f^-(0)$ &$-0.208$&$-0.184$&$-0.147$&$-0.174$&$-0.393$&$-0.322$&$18\%$\\ \hline
         $f^T(0)$&0.317&0.335&0.241&0.385&0.653&0.626&$-4.2\%$\\ \hline
    \end{tabular}
    \caption{Central value of LO and NLO form factors for $B\to K^*_0(1430)$ transitions. }
    \label{Kresult}
\end{table}

The main results of this analysis are the values of the form factors
$q^2=0$,
\begin{align}
   f^+(0)&=0.503
    \phantom{a}^{+0.052}_{-0.052}\Big|_{\rm twist\text{-}2}
    \phantom{a}^{+0.004}_{-0.004}\Big|_{\rm twist\text{-}3}
    \phantom{a}^{+0.056}_{-0.056}\Big|_{\bar f_{K_0^*}}
    \phantom{a}^{+0.020}_{-0.018}\Big|_{\overline{M},M},
    \notag\\
   f^-(0)&=-0.322
    \phantom{a}^{+0.051}_{-0.042}\Big|_{\rm twist\text{-}2}
    \phantom{a}^{+0.011}_{-0.011}\Big|_{\rm twist\text{-}3}
    \phantom{a}^{+0.036}_{-0.036}\Big|_{\bar f_{K_0^*}}
    \phantom{a}^{+0.019}_{-0.014}\Big|_{\overline{M},M},
    \notag\\
   f^T(0)&=0.626
    \phantom{a}^{+0.094}_{-0.094}\Big|_{\rm twist\text{-}2}
    \phantom{a}^{+0.009}_{-0.008}\Big|_{\rm twist\text{-}3}
    \phantom{a}^{+0.070}_{-0.070}\Big|_{\bar f_{K_0^*}}
    \phantom{a}^{+0.030}_{-0.023}\Big|_{\overline{M},M}.
\label{32}
\end{align}
Here, the individual uncertainties are obtained by varying the
corresponding input parameters within their quoted ranges while keeping
the remaining inputs fixed. The uncertainties associated with the
twist-2 contribution and the scalar decay constant $\bar f_{K_0^*}$ provide
the dominant contributions to the errors, whereas the
uncertainties associated with the twist-3 input and the Borel parameter
are comparatively smaller. The scalar decay constant and the Gegenbauer moments entering the different twist contributions are obtained from the same QCDSR analysis and are therefore expected to exhibit nontrivial correlations. In the absence of a covariance matrix that quantifies these
correlations, we vary the corresponding inputs independently, following
the procedure commonly adopted in the literature. This prescription
allows the individual sources of uncertainty to be identified
separately, although it may lead to a conservative, and potentially
slight overestimate, of the total uncertainty.

\begin{figure}[t]
    \centering

    % Top row
    \begin{subfigure}[b]{0.45\textwidth}
        \centering
        \includegraphics[width=\textwidth]{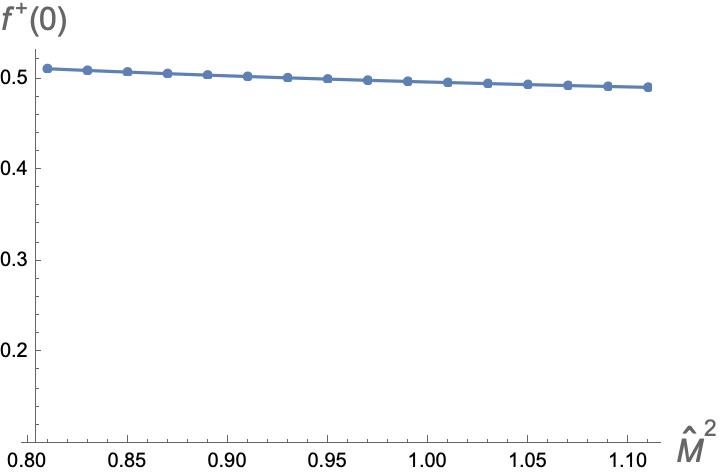}
        \caption{}
        \label{fig:fplusM}
    \end{subfigure}
    \hfill
    \begin{subfigure}[b]{0.45\textwidth}
        \centering
        \includegraphics[width=\textwidth]{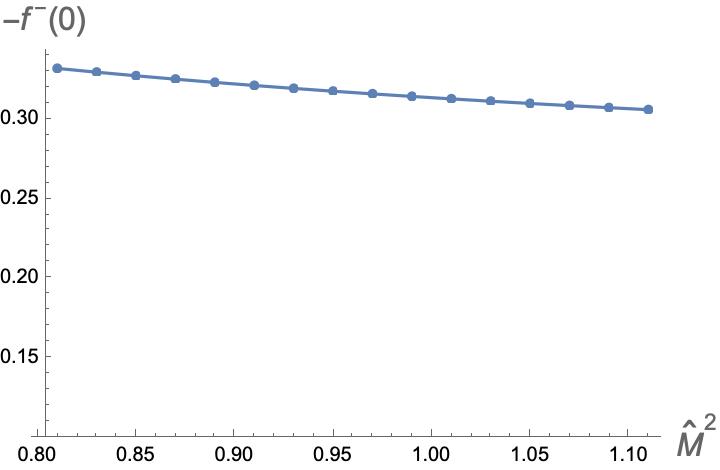}
        \caption{}
        \label{fig:fminusM}
    \end{subfigure}

    \vspace{0.5cm}

    % Bottom row
    \begin{subfigure}[b]{0.45\textwidth}
        \centering
        \includegraphics[width=\textwidth]{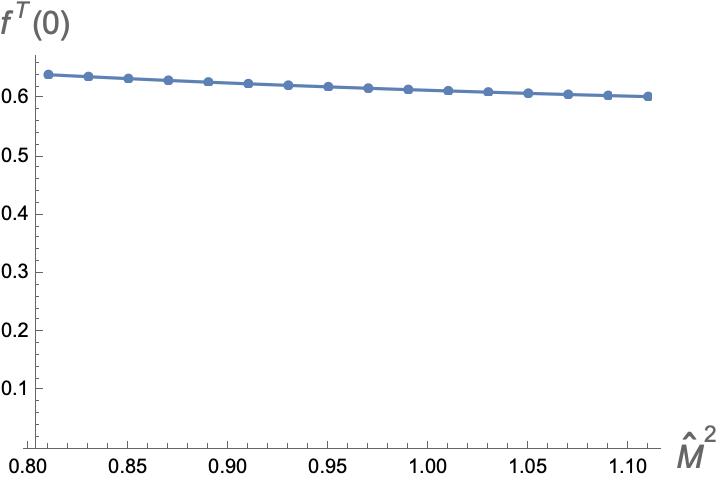}
        \caption{}
        \label{fig:fTM}
    \end{subfigure}

    \caption{Dependence of the $B\to K_0^*(1430)$ form factors on the
    Borel parameter $\hat{M}^2$ over the adopted Borel window $16-22$ $\mathrm{GeV^2}$. The LO and NLO
    results are shown by solid and dotted curves, respectively.}
    \label{fig:M2graphs}
\end{figure}

The dependence of the form factors on the Borel parameter $M^2$ over
the adopted Borel window is shown in Fig.~\ref{fig:M2graphs}. The form factors exhibit a relatively mild dependence on $M^2$ within the adopted window, indicating satisfactory stability of the sum rules. Form factor $f^-(r_1)$ is extracted from the combination $f^+(r_1)+f^-(r_1)$. The inclusion of the NLO corrections modifies the central values while preserving the overall Borel stability.

Table~\ref{comparison} presents our results for the $B\to K_0^*(1430)$
form factors together with their values obtained using other
approaches. Although these form factors have been studied using a
variety of methods in the literature, here we focus mainly on the
previous LCSR calculations that are most closely related to the present
analysis and span roughly the last two decades. In Ref.~\cite{Wang:2008da}, the form factors were calculated at LO using LCSR, including contributions up to twist-3. We can
reproduce their results at this order in the limit $m_S^2\to0$, which
corresponds to neglecting the corrections arising from the expansion
near the light cone. The small differences that remain can be
attributed mainly to the updated input parameters used in the present
analysis. Ref.~\cite{Wang:2014vra} extended this calculation by
including the $\mathcal{O}(\alpha_s)$ corrections to the twist-2
DA. In comparison, the present analysis includes
two additional effects: the corrections associated with the expansion
near the light cone and the $\mathcal{O}(\alpha_s)$ corrections to both
the twist-2 and twist-3 DAs. In this sense, our
calculation provides a more complete treatment of the LCSR at NLO compared to earlier work by \cite{Wang:2008da} at LO and twist-2 NLO \cite{Wang:2014vra}. The twist-3 NLO  corrections are not significant for $f^+(0)$ and is around 1\%. The difference with the Ref.~\cite{Wang:2014vra}  is mainly due to mass corrections. The twist-3 NLO corrections to  $f^-(0)$ and $f^T(0)$ compared to LO are 6\% and 15\%, respectively. Therefore, the addition of NLO twist-3 contributions along with light-cone expansion had been significant in our analysis.

%Despite the inclusion of the NLO twist-3 correction in the present calculation - the $f^+(0)$ form factor is numerically close to that of Ref.~\cite{Wang:2014vra} which has twist-2 NLO only.  This similarity, however, should not be interpreted as indicating that the NLO twist-3 correction is numerically unimportant. The corrections proportional to $m_S^2$ from the near-light-cone expansion give a negative contribution to the form factor and partially compensate the enhancement produced by the NLO twist-3 contribution. Consequently, the net change in $f^+(0)$ can appear relatively small. This compensation is less pronounced for $f^-(0)$, for which the $m_S^2$ corrections are comparatively smaller and the effect of the NLO twist-3 contribution is more visible. The difference between our result and that of Ref.~\cite{Wang:2014vra} is therefore more apparent in $f^-_{B^0\to K_0^*}(0)$.

The remaining results in Table~\ref{comparison} are included for
comparison but are obtained using different frameworks. In
Ref.~\cite{Aliev:2007rq}, the form factors were determined using
three-point QCD sum rules. Their results are considerably smaller than
those obtained in the analyses presented in Table \ref{comparison}. In Ref.~\cite{Han:2023pgf}, the form factors were
calculated at NLO using $B$-meson DAs within
soft-collinear effective theory (SCET). The QCD corrections were
matched onto $\mathrm{SCET}_{\mathrm{I}}$ operators at the hard scale
and subsequently onto $\mathrm{SCET}_{\mathrm{II}}$ operators at an
intermediate scale, with next-to-leading logarithmic (NLL) resummation
included. Finally, Ref.~\cite{Khosravi:2022fzo} employs $B$-meson
distribution amplitudes at leading order, including twist-5 and
three-particle contributions, and presents results using both the
exponential (Exp) and local-duality (LD) models
\cite{Lu:2018cfc}. The differences among these results reflect, in
part, the different sum-rule frameworks, non-perturbative inputs, and
treatments of higher-twist and radiative corrections.

%Since in this paper we found the twist-3 NLO corrections, it was instructive to compare the $\overline{\text{MS}-}$scheme results with that of the pole-mass scheme. We found a 4\% difference for the form-factor $f^+(0)$. The behavior on of NLO corrections though is different. In pole-scheme, there is a nice cancellation happening at high $q^2$ between the two twist-3 functions, i.e., $\phi_{3S}^{S}$ and $\phi_{3S}^{\sigma}$ while in \overline{\text{MS}}

\begin{table}[t]
    \centering
    \begin{tabular}{c c c c c}
    \hline\hline
         Sr.No&References&$f^+(0)$&$\qquad f^-(0)$&$\qquad f^T(0) $\vspace{0.1cm} \\ \hline
         \vspace{0.2cm}
        1& This work &$0.503^{+0.079}_{-0.078} $&$-0.322^{+0.066}_{-0.058}$&$0.626^{+0.121}_{-0.120}$\vspace{0.1cm} \\
        2&LCSR\cite{Wang:2008da}&$0.485^{+0.100}_{-0.100}$&$-0.412^{+0.102}_{-0.102}$&$0.60^{+0.140}_{-0.130}$\vspace{0.1cm}\\
        3&LCSR\cite{Wang:2014vra}&$0.523^{+0.070}_{-0.070}$&$-0.275^{+0.064}_{-0.064}$&$0.657^{+0.109}_{-0.109}$\vspace{0.1cm}\\
        
         4& QCDSR\cite{Aliev:2007rq}&$0.31^{+0.08}_{-0.08}$&$-0.31^{+0.07}_{-0.07}$& $-0.26^{+0.07}_{-0.07}$\vspace{0.1cm}\\
         
        5&LCSR \cite{Han:2023pgf}&$0.43$&$-0.42$&$0.58$\vspace{0.1cm}\\
 
        6&LCSR-Exp\cite{Khosravi:2022fzo}&$0.53^{+0.28}_{-0.22}$&$-0.51^{+0.13}_{-0.36}$&$0.72^{+0.39}_{-0.31}$\vspace{0.1cm}\\ 
        7&LCSR-LD\cite{Khosravi:2022fzo}&$0.60^{+0.34}_{-0.28}$&$-0.59^{+0.30}_{-0.41}$&$0.79^{+0.45}_{-0.35}$\vspace{0.1cm}\\  
 \hline\hline
    \end{tabular}
    \caption{Comparison of values of $B^0\to K_0^*(1430)$ form-factor calculated in this work with different approaches exist in the literature. The errors given in Eq. (\ref{32}) are combined using quadrature method. }
    \label{comparison}
\end{table}

\begin{figure}
    \centering
    \begin{subfigure}[b]{0.32\textwidth}
        \centering
        \includegraphics[width=\textwidth]{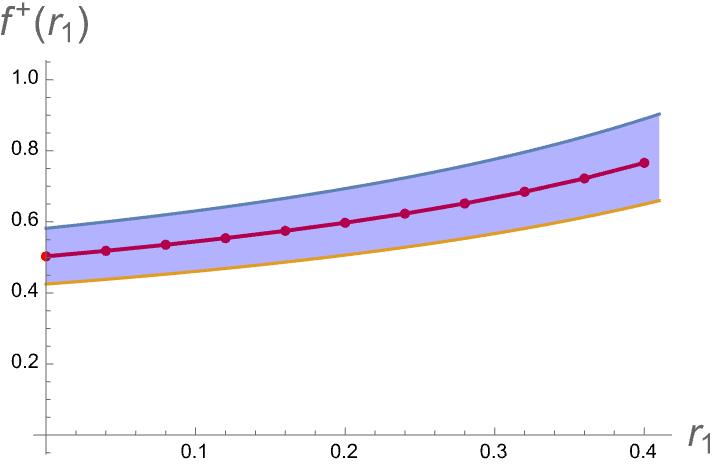}
    \end{subfigure}
    \hfill
    \begin{subfigure}[b]{0.32\textwidth}
        \centering
        \includegraphics[width=\textwidth]{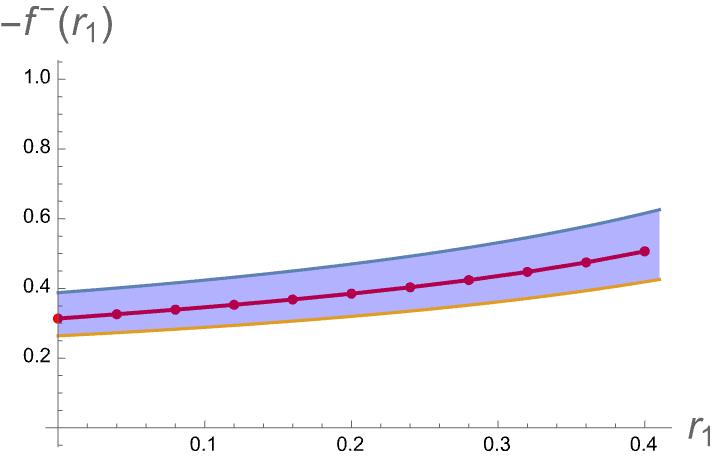}
    \end{subfigure}
    \hfill
    \begin{subfigure}[b]{0.32\textwidth}
        \centering
        \includegraphics[width=\textwidth]{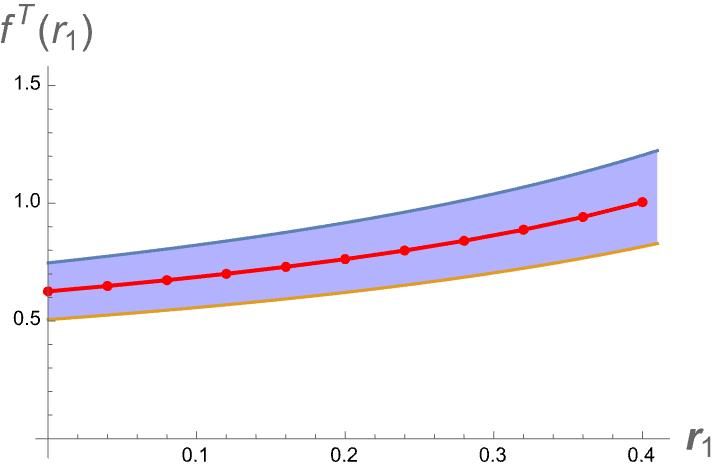}
    \end{subfigure}
    \hfill
    \caption{Form-factors' dependence upon $q^2\sim (0,8)\rm{GeV}^2$ is shown with corresponding uncertainties. }
    \label{fig:q2graphs}
\end{figure}

\section{Conclusion}\label{sec:conlcusion}
In this paper, the form-factors of $B-$meson to scalar mesons are calculated at NLO in LCSRs up to twist-3. The near-light cone expansion is used to accommodate the (slightly) massive scalar mesons to get generic expressions for the form-factor. A consistent NLO calculation is performed for $B\to K_0^*(1430)$ in $\overline{\text{MS}}-$scheme utilizing two daughter sum rules. The near-light cone expansion resulted in 5\% correction to LO form-factor $f^+(0)$ while 3\% for $f^T(0)$. The NLO corrections are important despite partial cancellation in $\tfrac{f_Bf_{BK^*_0}^{(+,T)}(0)}{f_B}$ using the LCSR for $B\to K_0^*(1430)$ and QCD sum rule for $f_B$ decay constant. The twist-3 radiative corrections for $f^+(0)$ is insignificant and is around 1\% while for $f^-(0)$ it is around 6\% . It is noted that twist-3 NLO corrections are especially significant for the tensor form-factor, $f^T$ for which it is 15\%, though when combined with the negative NLO contribution from twist-2 NLO, the total NLO contribution is $-4.2\%$. The major uncertainty in the form factors lie in the decay constant $\overline f_{K_0^*}$ and the twist-2 Gegenbaur moments. These two are correlated and a future experimental or lattice determination of the shape function will help ascertain their covariance and precise values when compared to the presented LCSR calculation.  
\appendix
\section{RG evolution of parameters}\label{App A-RG}

 The $b-$quark $\overline{\text{MS}}$-scheme mass from the renormalization group equation is: 
\begin{eqnarray}
    m_b(\mu)&=&m_b(m_b)\left(\frac{\alpha_s(\mu)}{\alpha_s(m_b)}\right)^{\frac{12}{13}}\notag\\
    \alpha_s(\mu)&=&\frac{1}{b_0 t}\left(1-\frac{b_1}{b_0^2}\frac{\log t}{t}+\frac{b_1^2  (\log^2 t-\log t-1)+b_0b_2}{b_0^4t^2}\right),\label{mbmuasmu}
\end{eqnarray}
where, the quantities $t=\log \frac{\mu^2}{\Lambda^2},\; b_0=\frac{33-2n_f}{12\pi},\; 
    b_1=\frac{153-19n_f}{24\pi^2},\;b_2=\frac{2857-\frac{5033}{9}n_f+\frac{325}{27}n_f^2}{128\pi^3},\;\Lambda=213\; \text{MeV}, 296\; \text{MeV}$ and $339\; \text{MeV}$ for flavors $n_f=5,4$ and 3, respectively. This yields the strong coupling values of $\alpha_s(3\;\text{GeV})=0.255$, and $\alpha_s(1\;\text{GeV})=0.484$. The scalar meson decay constant and the Gegenbaur moments corresponding to twist-2 and twist-3 DAs also evolve with scale and are determined via
    \begin{eqnarray}
        \overline{f}_S(\mu)&=&\overline{f}_S(\mu_0)\left(\frac{\alpha_s(\mu_0)}{\alpha_s(\mu)}\right)^\frac{4}{\beta},\quad\quad
        B_m(\mu)=B_m(\mu_0)\left(\frac{\alpha_s(\mu_0)}{\alpha_s(\mu)}\right)^{-\frac{\gamma_m + 4}{\beta}}\notag\\
        a_m(\mu)&=&a_m(\mu_0)\left(\frac{\alpha_s(\mu_0)}{\alpha_s(\mu)}\right)^{-\frac{\gamma_S}{\beta}},\quad
        b_m(\mu)=b_m(\mu_0)\left(\frac{\alpha_s(\mu_0)}{\alpha_s(\mu)}\right)^{-\frac{\gamma_T}{\beta}}\label{RGmoments}
    \end{eqnarray}
where $\beta=11-2n_f/3$, and the anomalous dimensions are
\begin{eqnarray*}
    \gamma_m&=& C_F\left(1-\frac{2}{(m+1)(m+2)}+4\sum_{j=2}^{m+1}\frac{1}{j}\right)\;,\quad\quad
    \gamma_S= C_F\left(1-\frac{8}{(m+1)(m+2)}+4\sum_{j=2}^{m+1}\frac{1}{j}\right)\notag\\
    \gamma_T&=& C_F\left(1+4\sum_{j=2}^{m+1}\frac{1}{j}\right)
\end{eqnarray*}
\section{Daughter sum rule for threshold parameter}\label{App B-daugter SR}
After applying quark–hadron duality and performing the Borel transformation with respect to the momentum variable $(P+q)^2$, the LCSR for the $B\to K^*_0\left(1430\right)$ form factor can be written schematically as
\begin{equation}
    f_B\, f^+(q^2)\,
e^{-m_B^2/M^2}
=
\Pi_{\rm }( M^2, s_0^B,q^2),
\end{equation}
where $\Pi_{\rm }( M^2, s_0^B,q^2)$ denotes the Borel-transformed correlation function with the hadronic continuum above the effective threshold \(s_0^B\) subtracted. Explicitly, it has the form
\begin{equation}
    \Pi( M^2,s_0^B,q^2)
=
\int_{m_b^2}^{ s_0^B} ds\,
\rho^{\rm QCD}(s,q^2)\,
e^{-s/M^2},
\end{equation}
where \(\rho^{\rm QCD}(s,q^2)\) contains the perturbative spectral density, including the LO and NLO contributions.

The parameter \(s_0^B\) is not known a priori and is therefore constrained by requiring that the sum rule reproduce the physical \(B\)-meson mass, i.e., %This is achieved by taking the logarithmic derivative of the sum rule with respect to the inverse Borel parameter,
\begin{eqnarray}
    m_B^2 &=&  -\frac{d}{d(1/M^2)}\ln \Pi(M^2,s_0,q^2),\nonumber\\
    &=&-\bigg[\frac{1}{\Pi(M^2,s_0^B,q^2)}\bigg]\frac{d}{d(1/M^2)} \Pi(M^2,s_0,q^2)
\end{eqnarray}
one obtains the daughter sum rule
\begin{equation}
m_B^2=\frac{\displaystyle\int_{m_b^2}^{s_0^B} ds\,s\,\rho(s,q^2)e^{-s/M^2}}{\displaystyle\int_{m_b^2}^{s_0^B} ds\,\rho(s,q^2)e^{-s/M^2}}
\end{equation}

Thus the daughter sum rule provides a direct test of whether the chosen effective threshold and Borel parameter give a physically consistent description of the \(B\)-meson pole.

In practice, one may determine an effective Borel-dependent continuum threshold as:
$$
s_0^B=s_0^B\left(M^2\right)\;,
$$
by imposing
\begin{equation}
    m_B^2(M^2,s_0^B)=m_{B,\mathrm{phys}}^2.
\end{equation}
The resulting $s_0^B\left(M^2\right)$ is then used in the original sum rule for the form factor.

\section{Two-point sum rule for $f_B$}\label{App C-daugter SR}

The general two-point QCD sum rule expression for the \(B\)-meson decay constant \(f_B\), after the Borel transformation and continuum subtraction, can be written as
\begin{eqnarray}
    m_B^4 f_B^2 e^{-m_B^2/\overline{M}^2}&=&\int_{m_b^2}^{\bar s_0^B} ds \rho_B(s,\mu)+\Pi_{B}^{\text{power}}(\overline{M}^2,\mu)
\end{eqnarray}
where $\Pi_{B}^{\text{power}}$ contains power/condensate corrections. The spectral density $\rho_B(s,\mu)=\rho_B^{\text{LO}}(s,\mu)+\rho_B^{\text{NLO}}(s,\mu)$ up to $\mathcal{O}(\alpha_s)$ \cite{Jamin:2001fw} is given as
\begin{eqnarray}
    \rho_B^{\text{LO}}&=&\frac{3 (s-m_b^2)^2}{8 \pi^2 s}\notag\\
    \rho_B^{\text{NLO}}&=&\frac{\alpha_s C_F}{\pi}\frac{3 (m_b^2)}{8 \pi^2}\frac{s}{2}(1-x)
\Bigg\{(1-x)\Bigg[4\operatorname{Li}_2(x)+2\ln x\ln(1-x)-(5-2x)\ln(1-x)
+(1-2x)(3-x)\ln x\Bigg.\Bigg.\notag\\
&&\Bigg.\Bigg. +3(1-3x)\ln\left(\frac{\mu^2}{m_b^2}\right)+\frac{1}{2}(17-33x)\Bigg]\Bigg\}\;, 
\end{eqnarray}
where $x=\frac{m_b^2}{s}$. The dimension-3 quark condensate contribution (including $\mathcal{O}(\alpha_s)$	correction and the mixed quark-gluon condensate correction through $(m_0^2)$), dimension-4 gluon condensate contribution and dimension-6 four quark contribution are given as 
\begin{eqnarray}
    \Pi_{B}^{\text{power}}&=&-m_b\langle\bar q q\rangle\left(1+\frac{\alpha_s C_F}{\pi}\delta_1(\overline{M}^2,m_b^2)
+\frac{m_0^2}{2\overline{M}^2}\left(1-\frac{m_b^2}{2\overline{M}^2}
\right)\right)\nonumber\\
&&\qquad+\frac{1}{12}\left\langle\frac{\alpha_s}{\pi}GG\right\rangle
-\frac{16\pi\alpha_s\langle\bar q q\rangle^2}{27\overline{M}^2}
\left(1-\frac{m_b^2}{4\overline{M}^2}-\frac{m_b^4}{12\overline{M}^4}\right)
\Bigg]\;,
\end{eqnarray}
where the quark-condensate density at $\mathcal{O}(\alpha_s)$ is given by the function
\begin{align}
\delta_1(\overline{M}^2,m_b^2)
={}&
-\frac{3}{2}\Bigg[\Gamma\left(0,\frac{m_b^2}{\overline{M}^2}\right)
e^{m_b^2/\overline{M}^2}-1-\left(1-\frac{m_b^2}{\overline{M}^2}\right)
\left(\ln\left(\frac{\mu^2}{m_b^2}\right)+\frac{4}{3}\right)\Bigg]\;.
\label{eq:delta1}
\end{align}
Here $\Gamma(n,z)$ is the incomplete gamma function. The $\overline{M}$ and $\overline{s}_0^B$ are, respectively, the Borel parameter
and the effective threshold of the 2pSR. They can be determined similar to the procedure laid down for the $B\to K_0^*\left(1430\right)$ LCSR in Appendix \ref{App B-daugter SR}  by requiring 
\begin{equation}
m_B^2(\overline{M}^2,\overline{s}_0^B)=m_{B,\mathrm{phys}}^2\;.
\end{equation}

\section*{Acknowledgement}
The author (Arslan Sikandar) would like to express deep gratitude to  Bla\v{z}enka Meli\'c for encouraging using  light-cone sum rules for $B-$ decays and supported his stay at Ru\dj er Boskovi\'c Institute, Zagreb, Croatia. He also likes to express special thanks to the Mainz Institute for Theoretical Physics (MITP) of the Cluster of Excellence PRISMA$^+$ (Project ID 390831469), for its hospitality and support. The author had fruitful discussion at MITP with Danny Van Dyk on the use of daughter sum rule for consistent calculation.

\end{document}